\documentclass{iau307}
\usepackage{graphicx}
\usepackage{natbib}
\usepackage{url}
\bibpunct{(}{)}{;}{a}{}{,}

\title[Stability boundaries in the sHR diagram] 
{Stability boundaries for massive stars in the sHR diagram}

\author[H. Saio, C. Georgy, G. Meynet]   
{Hideyuki Saio$^1$, Cyril Georgy$^2$,
 \and Georges Meynet$^3$}

\affiliation{$^1$Tohoku University, Japan \\ email: {\tt saio@astr.tohoku.ac.jp} \\[\affilskip]
$^2$Keele University, UK, ~
$^3$Geneva observatory, Switzerland}

\pubyear{2014}
\volume{307} 
\pagerange{}
\jname{New windows on massive stars: asteroseismology, interferometry, and spectropolarimetry}
\editors{G. Meynet, C. Georgy, J.H. Groh \& Ph. Stee, eds.}

\begin{document}

\maketitle

\begin{abstract}
Stability boundaries of radial pulsations in massive stars are compared with 
positions of variable and non-variable blue-supergiants in the spectroscopic 
HR (sHR) diagram \citep{lan14}, whose vertical axis is 
$4\log T_{\rm eff}-\log g \,(= \log L/M)$. 
Observational data indicate that variables tend to have higher
 $L/M$ than non-variables in agreement with the 
theoretical prediction. However, many variable blue-supergiants are found to 
have values of  $L/M$ below the theoretical stability boundary;
i.e., surface gravities seem to be too high by around 0.2-0.3 dex.

\end{abstract}

\firstsection 
\vskip 0.5cm
Recently, a spectroscopic HR (sHR) diagram was introduced by \citet{lan14}. 
Its horizontal axis is  $\log T_{\rm eff}$, common to the ordinary HR 
diagram, while the vertical axis is $4\log T_{\rm eff} - \log g$ (in solar units). 
One of the differences from the ordinary HRD is that no distance information 
is needed for plotting stars in the sHRD, although accurate estimates of 
the surface gravity $\log g$ are needed.
Theoretically, the vertical axis is equal to $\log(L/M)$ (in solar units), 
which is affected by mass loss 
in the post-main-sequence evolution more sensitively than 
$\log L$ in the ordinary HRD.
Because of  these properties, sHRD is useful,
in addition for the properties discussed by \citet{lan14},
for comparing pulsational stability boundaries with observed 
positions of blue supergiants (BSG).

\vskip 0.3cm
A massive star becomes a BSG (BSG1) after the main-sequence evolution,
 and evolves to the red supergiant (RSG) stage. 
After losing significant mass in the RSG stage, the star becomes BSG again (BSG2)
\citep{eks12}.
\citet{sai13} found that radial pulsations in the BSG region are excited
(by the strange mode instability) 
only in BSG2 but not in BSG1 (Fig. 1).
Since the luminosity is not very different 
between BSG1 and BSG2 for a given mass,
the distributions of BSG1s and BSG2s, and hence the variable and non-variable BSGs 
in the HR diagram are similar and mixed.
In the sHRD diagram, however, the evolutionary track is affected more strongly
by mass loss, and  BSG2 is located significantly 
higher than BSG1, so that  the pulsational instability region, which is governed mainly
by $L/M$,
is separated roughly irrespective to the mass (right panel of Fig.1).

\begin{figure}[!h]
\begin{center}
\includegraphics[width=0.49\textwidth]{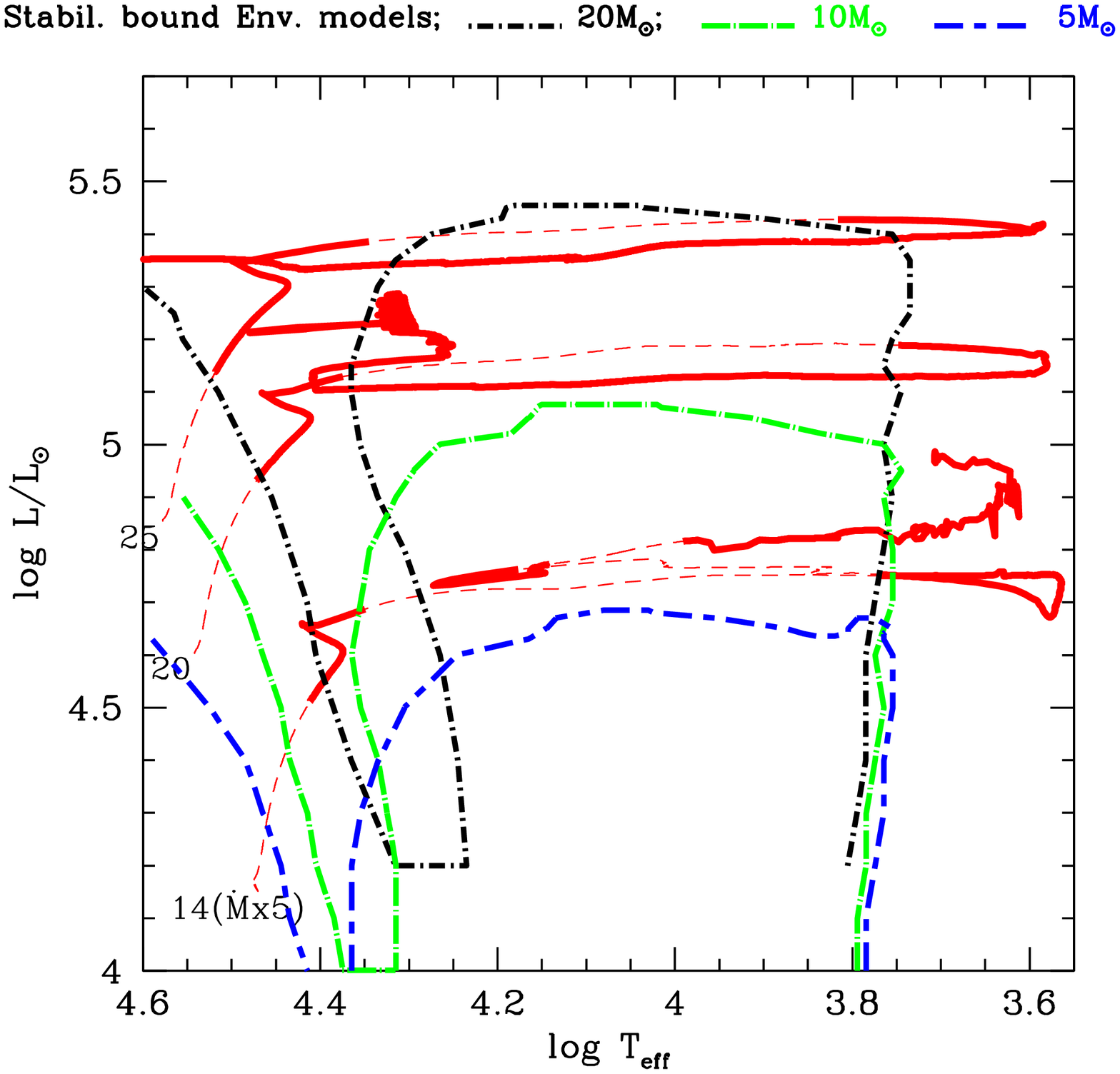} 
\includegraphics[width=0.49\textwidth]{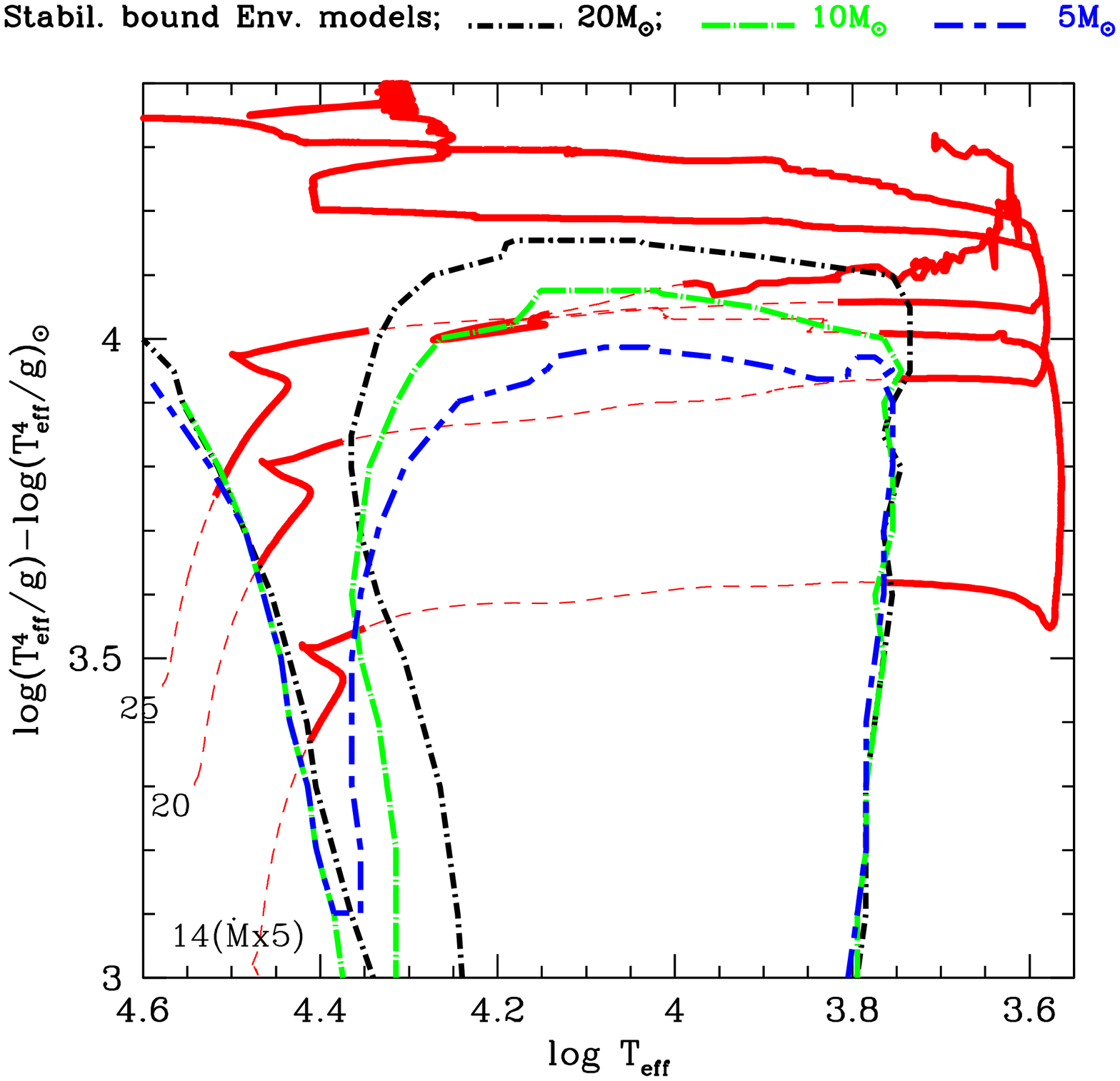} 
\caption{Stability boundaries of radial pulsations for envelope models of
$20\,M_\odot, 10\,M_\odot$, and $5\,M_\odot$ are shown in the ordinary
HR diagram (left panel) and sHR diagram (right panel). 
Parts of  thick solid lines in evolutionary tracks indicate where
at least one radial modes are excited. 
}
\label{fig1}
\end{center}
\end{figure}

\begin{figure}[!h]
\begin{center}
\includegraphics[width=0.49\textwidth]{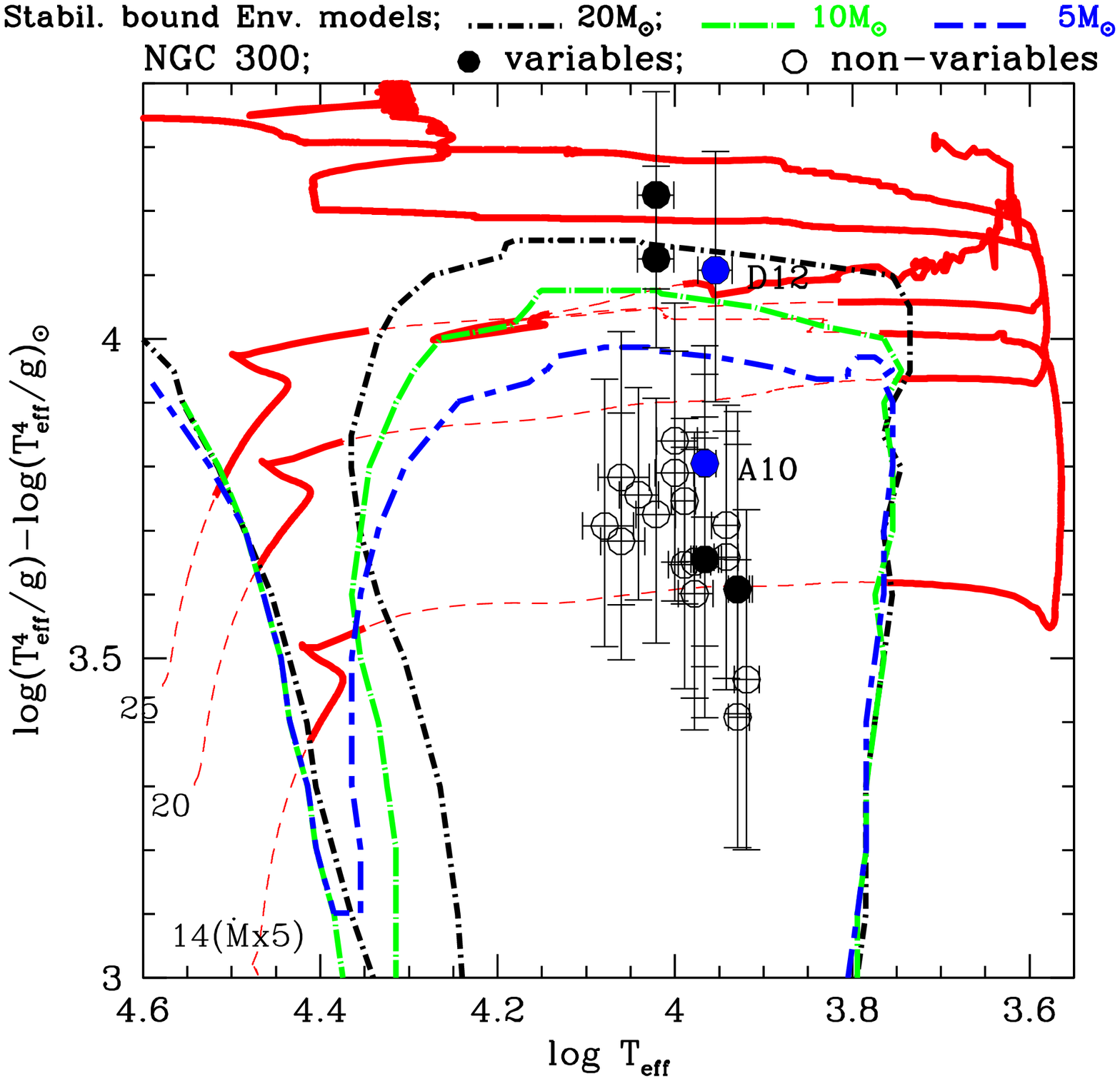} 
\includegraphics[width=0.49\textwidth]{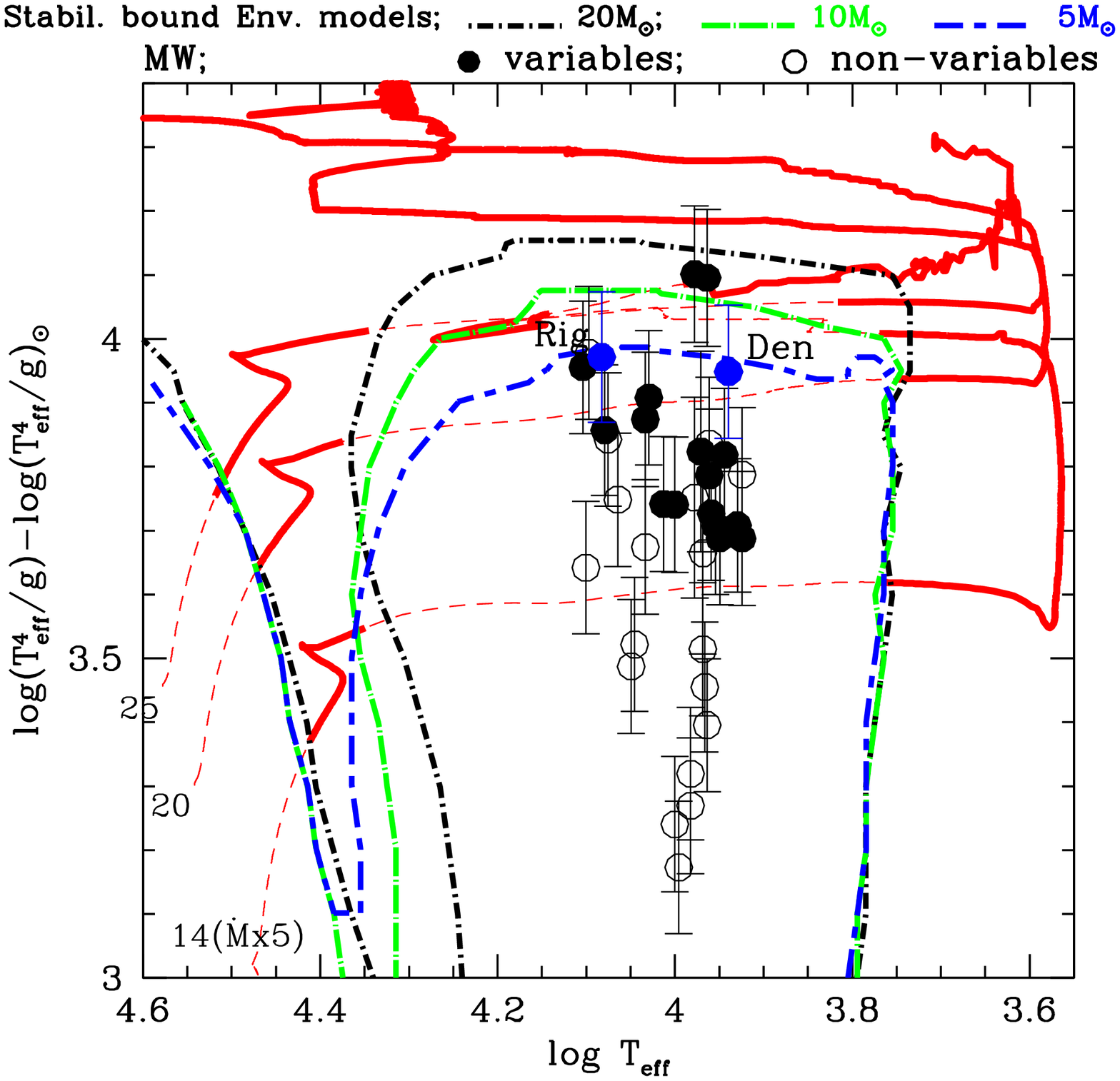} 
\caption{BSGs
in NGC\,300 \citep[left panel;][]{bre04,kud08} and BSGs in the 
Milky Way \citep[right panel;][]{fir12} are plotted in 
the sHR diagram.  Filled (open) circles indicate variable (non-variable) BSGs.
Stability boundaries shown in Fig.\,1 (right panel) are also shown for comparison.
The stars D12 and A10 in NGC\,300 (left panel) show regular light curves 
which look consistent 
with radial pulsations.
In the right panel `Rig' and `Den' stand for Rigel and Deneb, respectively.
}
\label{fig2}
\end{center}
\end{figure}

Fig.\,\ref{fig2} compares BSGs in NGC\,300 (left panel) 
and BSGs in the Milky Way (right panel) with theoretical stability boundaries
in the sHRD.
Generally, variable BSGs are located above the non-variable BSGs 
(with some exceptions) consistent with the 
theoretical prediction.
However, observational non-variable/variable boundary (for MW) seems lower than 
the theoretical stability boundary of pulsation; i.e., observational values of $\log g$ seem
somewhat too high. 
The reason for the discrepancy is not clear at present. 

\bibliographystyle{iau307}
\bibliography{Biblio_p37}

\end{document}